\documentclass[twocolumn,preprintnumbers,aps,pra,showpacs,amsmath,amssymb,superscriptaddress]{revtex4-1}
\usepackage{hyperref} 
\usepackage{adjustbox}
\usepackage{subcaption}
\usepackage{graphicx}
\usepackage{color}
\usepackage{bm}
\usepackage{amsmath}
\usepackage{mathtools}
\usepackage{epstopdf}
\begin{document}

\title{Thresholded Quantum LIDAR \--- Exploiting Photon-Number-Resolving Detection }

\author{Lior Cohen} 
\affiliation{%
Hearne Institute for Theoretical Physics, and Department of Physics and Astronomy, Louisiana State University, Baton Rouge, Louisiana 70803, USA.
}
\affiliation{%
cohen1@lsu.edu}
\author{Elisha S. Matekole} 
\affiliation{%
Hearne Institute for Theoretical Physics, and Department of Physics and Astronomy, Louisiana State University, Baton Rouge, Louisiana 70803, USA.
}
\author{Yoni Sher}
\affiliation{%
School of Computer Science and Engineering, Hebrew University of Jerusalem, Jerusalem 91904, Israel.}
\author{Daniel Istrati}
\affiliation{%
Racah Institue for Physics, Hebrew     University of Jerusalem, Jerusalem          91904, Israel.}
\author{Hagai S. Eisenberg}
\affiliation{%
Racah Institue for Physics, Hebrew     University of Jerusalem, Jerusalem          91904, Israel.}
\author{Jonathan P. Dowling}
\affiliation{%
Hearne Institute for Theoretical Physics, and Department of Physics and Astronomy, Louisiana State University, Baton Rouge, Louisiana 70803, USA.
}
\affiliation{%
	NYU-ECNU Institute of Physics at NYU Shanghai, 3663 Zhongshan Road North, Shanghai, 200062, China.
	}
\affiliation{%
LCAS-Alibaba Quantum Computing        Laboratory, CAS Center for Excellence in     Quantum Information and Quantum Physics,     University of Science and Technology of     China, Shanghai 201315, China.
    }
\affiliation{%
National Institute of Information and Communications Technology,
4-2-1, Nukui-Kitamachi, Koganei, Tokyo 184-8795, Japan
    }

\date{\today}



\begin{abstract}

We present a technique that improves the signal-to-noise-ratio (SNR) of range-finding, sensing, and other light-detection applications.
The technique filters out low photon numbers using photon-number-resolving detectors (PNRDs). This technique has no classical analog and cannot be done with classical detectors. 
We investigate the properties of our technique and show under what conditions the scheme surpasses the classical SNR. Finally, we simulate the operation of a rangefinder, showing improvement with a low number of signal samplings and confirming the theory with a high number of signal samplings. 

\end{abstract}

\flushbottom
\maketitle
\thispagestyle{empty}

\textit{Introduction.}\----Electromagnetic radiation is regularly used for measuring and sensing the physical world. 
One particular sensing method, namely, laser range-finding and Light Detection and Ranging (LIDAR) is under continuous development. Increasing the range requires sensitive detectors, and more recently, single-photon detectors (SPDs) \cite{warburton2007subcentimeter,howland2013photon,pawlikowska2017single,li2019single}, and photon-number-resolving detectors (PNRDs)  \cite{bao2014laser,sher2018low} have been used for this purpose. 

It is an ongoing question what quantum optics can contribute to applications like LIDAR. 
It has been proven that loss, such as in rangefinders and LIDARs, eliminates most quantum effects \cite{dorner2009optimal,lee2009optimization}, thus, it is ineffective to use quantum states of light for those applications, rather than classical light such as coherent states \cite{jiang2013super}.
However, many proven quantum effects are not a result of using quantum states, but of using  quantum detection of these states. For example, Bell-inequality violations are commonly attributed to the use of entangled states \cite{qian2015shifting}. However, all-optical demonstrations have been done with Gaussian states, such as spontaneous parametric down-conversion \cite{giustina2013bell}. 
It is well known that Bell's inequalities are satisfied when both the state and the detection are Gaussian \cite{bell1987speakable}, thus, in all-optical demonstrations, Bell-inequality violations are caused by the non-Gaussian single-photon detection \cite{giustina2013bell}.
Having said that, even though rangefinders and LIDARs are operated with coherent states, quantum detection strategies such as parity \cite{jiang2013super}, and photon thresholding (filtering out low photon-numbers) \cite{bao2014laser} might still give a quantum advantage.
In this paper, we rigorously derive the SNR improvement of threshold detection over intensity detection. 

\begin{figure}[b]
    \includegraphics[width=.5\textwidth]{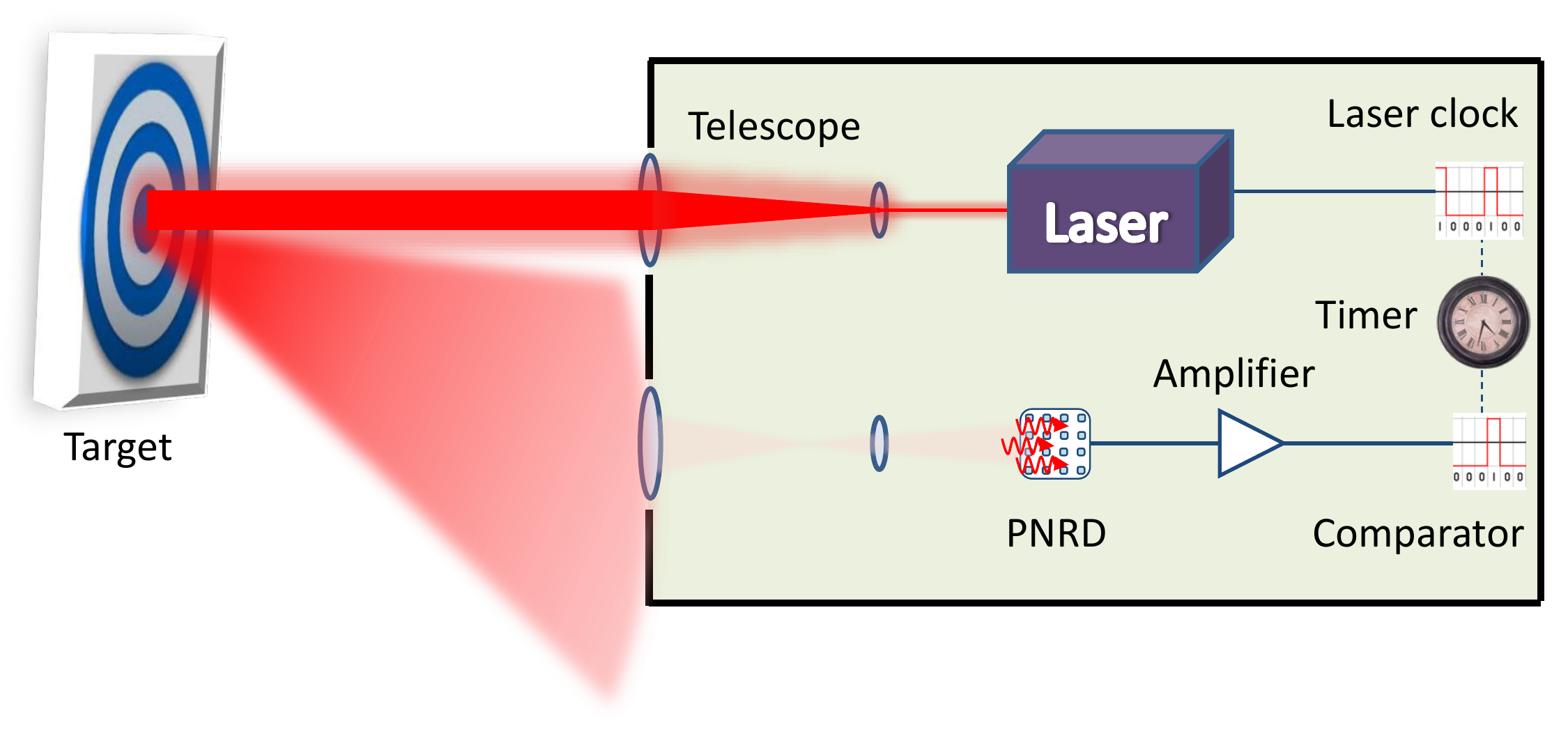}
    \caption{Illustration of the rangefinder system. A laser pulse is sent to a remote target and a small portion is reflected back into the device. After spatial and spectral filtering, the light is detected by a PNRD. Then, the photon number is thresholded by thresholding the voltage height. A one-bit comparator stops the timer when a voltage peak, caused by the detection of a bunch of photons, exceeds the voltage threshold. }
    \label{fig:system}
\end{figure}

One form of laser range-finding is illustrated in Fig. \ref{fig:system}. By sending short pulses of light, and recording their return time, one can measure the range to a target using the speed of light. The range-finding information can be extended to three-dimensional imaging by adding spatial resolution to the detection. Spatial resolution can be obtained by a gated camera \cite{busck2004gated}, raster scanning \cite{pawlikowska2017single} or blocking masks \cite{howland2013photon,sher2018low}. The last method also provides compressed data acquisition, where the number of required measurements is far less than the number of image pixels.  

In daylight range-finding, the classical noise from solar radiation dominates the quantum noise, the latter of which is due to the photon-number fluctuations of the coherent source.
Solar radiation is a blackbody radiation, and thus, single-mode sunlight has thermal photon-statistics:

\begin{equation}
    p_{\rm th}(n) = \frac{\bar{n}_{\rm th}^n}{(\bar{n}_{\rm th}+1)^{n+1}}\,,
\end{equation}
where $p_{\rm th}(n)$ is the probability to measure \textit{n}-photons within the coherence time, and $\bar{n}_{\rm th} = (e^{\hbar\omega/k_BT}-1)^{-1}$ is the average photon number, $\hbar$ and $k_B$ are the Dirac and Boltzmann constants and $\omega$ is the light frequency. 
The laser is a coherent light source and thus has a Poisson photon distribution:

\begin{equation}
    p_{\rm p}(n) = e^{-\bar{n}_{\rm p}} \frac{\bar{n}_{\rm p}^n}{n!}\,,
\end{equation}
where $\bar{n}_{\rm p}$ is the average photon number.
Since the solar flux is continuous, identifying the signal is equivalent to distinguishing a mixture of coherent and thermal light from thermal light alone. The mixture has mixed photon-statistics \cite{dovrat2012measurements}, $p(n) = \sum_{m=0}^n p_{\rm p}(m)p_{\rm th}(n-m)$ which can be written as

\begin{equation}
    p(n) = e^{\frac{\bar{n}_{\rm p}}{x}-\bar{n}_{\rm p}} \frac{x^n}{n!}  \Gamma\Big(\frac{\bar{n}_{\rm p}}{x},n+1\Big)\,,
\end{equation}
where $x = {\bar{n}_{\rm th}}/{(\bar{n}_{\rm th}+1)}$, and $\Gamma(y,n+1) = n!e^{-y}\sum_{m=0}^n ({y^m}/{m!}) = \int_y^\infty t^n e^{-t}dt $ is the incomplete gamma function.

\textit{Quantum SNR versus classical SNR.}\----Typically, in quantum sensing technologies, it is the shot-noise limit (SNL) that is beaten \cite{aasi2013enhanced,israel2014supersensitive}. While sub-SNL sensitivity can be obtained when the classical noise is negligible, it is a much harder task when the classical noise is dominant \cite{escher2011general,cohen2016demonstration}. Nevertheless we show that even in this regime, the SNR of quantum detection schemes can still surpass the SNR of classical detection schemes. 

Let us compare the classical intensity and our quantum-thresholding detection. Here the signal is regarded as the detection output with the coherent light, and the noise with the thermal light alone. As standard intensity detection is sensitive only to the average number of detected photons, the average photon number of the thermal light alone is the noise and the sum of the average photon-number of the two light sources is the signal. Thus, the classical SNR is

\begin{equation}
    {\rm SNR_c} = \frac{\bar{n}_{\rm p}+\bar{n}_{\rm th}} {\bar{n}_{\rm th}}\,.
\end{equation}

Threshold detection has a binary outcome; it is zero \---- if the detected photon number is below the threshold photon number, and one \---- if the detected photon number is above the threshold photon number. 
The signal of threshold detection is proportional to the probability of successfully exceeding the threshold when coherent light also hits the detector. The noise is proportional to the probability of exceeding the threshold when only thermal light hits the detector. These probabilities are calculated by summing all the photon-number statistics above $N$, the threshold photon-number. 

Thus, the noise is $\nu \sum_{n=N}^\infty p_{\rm th}(n) = \nu x^N$, and the signal  $\nu\sum_{n=N}^\infty p(n) = \nu\big[1 - \sum_{n=0}^{N-1} p(n)\big]$, where $\nu$ is the number of experimental repetitions. After substituting $p(n)$, reordering the sums and summing over \textit{n}, we are left with, $\nu\big[1 - \sum_{m=0}^{N-1} \big(1-x^{N-m}\big)p_{\rm p}(m)\big]$. Using the formula of the incomplete gamma function and dividing by the noise, we get that the SNR for threshold detection is:

\begin{equation}
    {\rm SNR_q} = \frac{1-\Big(\frac{\Gamma(\bar{n}_{\rm p},N)}{(N-1)!} - \frac{\Gamma(\frac{\bar{n}_{\rm p}}{x},N)}{(N-1)!} e^{\frac{\bar{n}_{\rm p}}{x}-\bar{n}_{\rm p}} x^N\Big)} {x^N}\label{Eq:quSNR}\,.
\end{equation}
Notice that the noise exponentially decays with the threshold number. This decay eventually gives the SNR improvement that we will see in the following.   

We wish to get some insights into the expression of Eq. \ref{Eq:quSNR}. 
First, we differentiate the SNR with respect to $\bar{n}_{\rm p}$, 
\begin{equation*}
\frac{\partial}{\partial\bar{n}_{\rm p}}{\rm SNR_q} = \big(\frac{1}{x}-1\big) \frac{\Gamma(\frac{\bar{n}_{\rm p}}{x},N)}{(N-1)!} e^{\frac{\bar{n}_{\rm p}}{x}-\bar{n}_{\rm p}} x^N>0\,    
\end{equation*}
which means that the SNR is a monotonically increasing function of the coherent mean-photon number regardless of the threshold and averaged thermal photon-number. This dependence is expected since increasing the signal intensity should increase the SNR. 

Next, we check the threshold dependence on photon number. The difference $[{\rm SNR_q}(N+1)-{\rm SNR_q}(N)]$ can be written as $[\sum_{n=N}^\infty p(n+1) - \sum_{n=N}^\infty p(n)x]/{x^{N+1}}$, where the first summation is transformed as $n \rightarrow  n+1$. Now the two summations can be regrouped into one, and its argument is $(1-x)p_{\rm p}(n+1)$. Thus, the SNR obeys 
\begin{equation}
    [{\rm SNR_q}(N+1)-{\rm SNR_q}(N)] = \frac{1-x}{x^{N+1}}\sum_{n=N}^\infty p_{\rm p}(n+1)>0\,,\label{Eq:therMonoton}
\end{equation}
i.e, taking larger photon-number thresholds increases the SNR for any intensity of the coherent and thermal light.

\begin{figure}[tb]
\centering
\includegraphics[width=\linewidth]{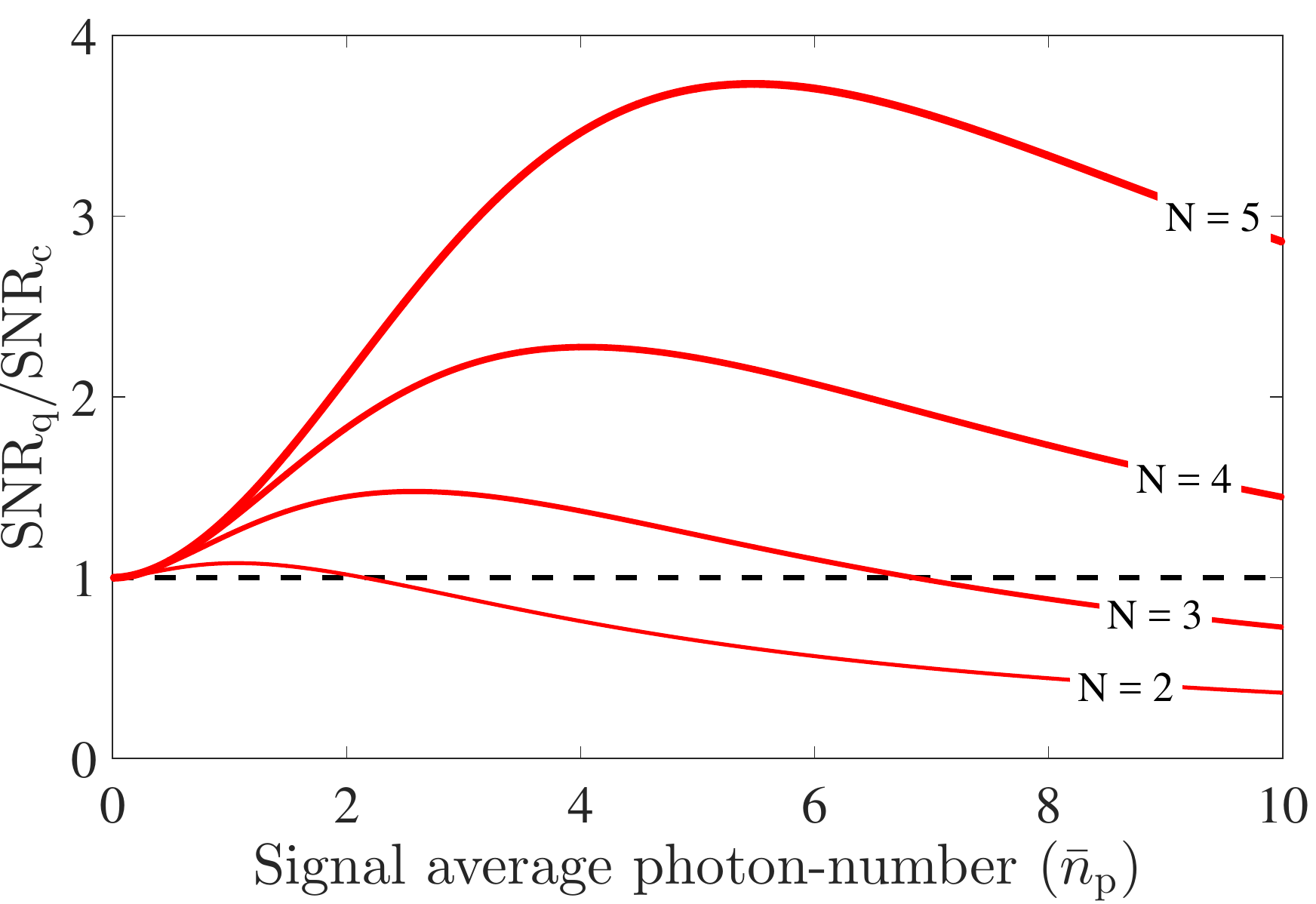}
\caption{The ratio of the quantum and classical SNR for fixed thermal average photon-number of one. Thresholds of $N=2,3,4,5$ are plotted where a thicker line corresponds to a higher threshold. The dashed black line at one represents the limit, above which the quantum scheme gets a better SNR.}
\label{fig2}
\end{figure}

In order to demonstrate the advantage of our quantum scheme, Fig. \ref{fig2} shows the ratio of the quantum and classical SNR for a fixed averaged-thermal photon number of one. Different threshold photon numbers are plotted with different line widths.

\textit{Discussion}.\----For many average signal and threshold photon numbers, the ratio of SNR is above one, which means that the quantum SNR exceeds the classical SNR. This improvement is a result of the difference between the signal and noise photon distribution. The thermal distribution is dominant near the low photon numbers, whereas the Poisson distribution is more dominant near the mean photon number (see Fig. S1 in the  supplementary material \cite{WinNT}). By using threshold detection we exclude low photon numbers where the noise is dominant. 

As shown in Eq. \ref{Eq:therMonoton}, the quantum SNR increases when a larger photon number threshold is used. Thus, the ratio of the two SNRs increases with the threshold, since the classical SNR is independent of the threshold.
However, taking threshold much larger than the average photon number will cause substantial decrease in the successful threshold detection. Any practical application should choose the threshold photon number in accordance with this trade-off; higher threshold means higher SNR but lower successful threshold detection, lower threshold means higher successful threshold detection but lower SNR. For a practical rangefinder or LIDAR, threshold detection success should be every couple of trials. Thus, in the regime of a few detected signal photons, the best improvement is around four.
\begin{figure}[tb]
\centering
\begin{subfigure}[b]{1\columnwidth}
   \includegraphics[angle=0,width=1\columnwidth]{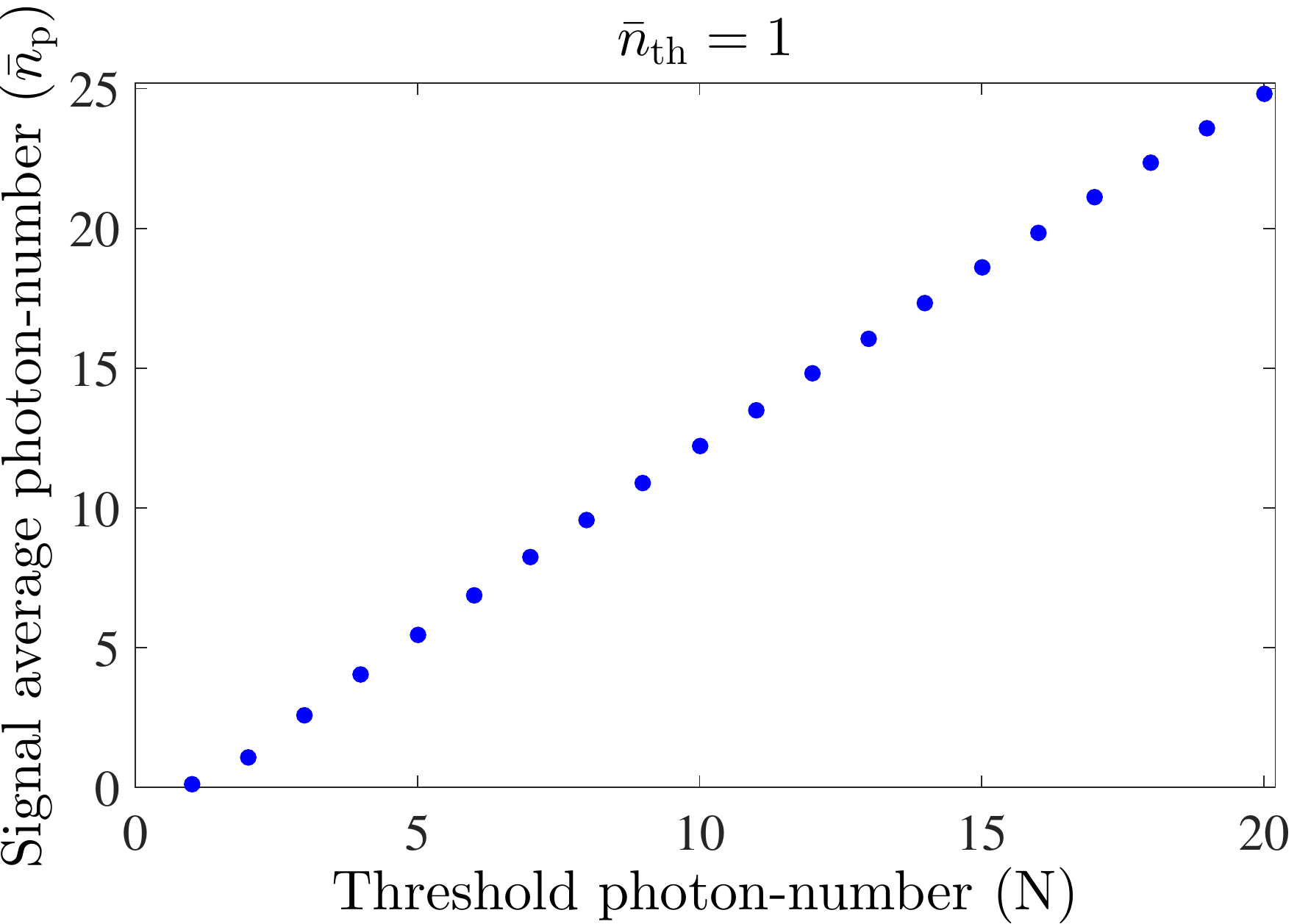}
   \caption{}
   \label{fig:3a}
\end{subfigure}
\begin{subfigure}[b]{1\columnwidth}
   \includegraphics[angle=0,width=1\columnwidth]{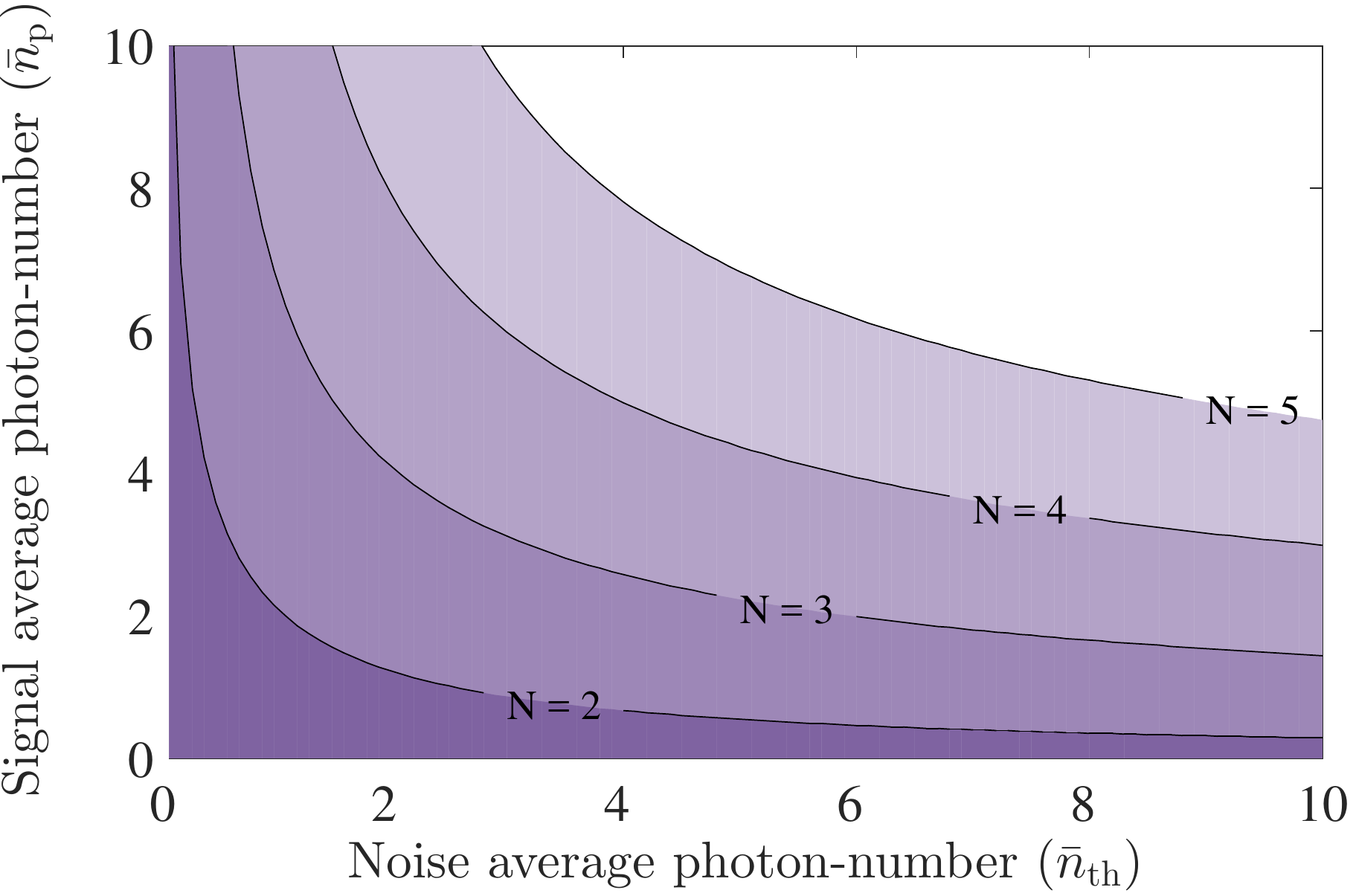}
   \caption{}
   \label{fig:3b}
\end{subfigure}
\caption{\textbf{(a)} The coherent light (signal) intensity that achieve the best improvement with respect to the classical detection scheme for  fixed thermal average photon-number of one. \textbf{(b)} Parameter-space representation of the quantum improvement. The line denotes the limit of quantum improvement, where below the line the threshold detection gives higher SNR than the classical detection, for particular threshold number, N. The area under the line increases for larger threshold numbers, showing the improvement achieved by taking larger threshold.} \label{fig3}
\end{figure}

In Fig. \ref{fig2}, for every threshold there is an averaged signal photon number where the improvement is maximal. In Fig. \ref{fig:3a} this maximum mean photon number is plotted as a function of the threshold. The improvement is maximal where the threshold is around the mean photon number. This observation can be understood by the fact that the coherent light has a more localized distribution than the thermal light, i.e. the variance of Poisson distribution equals the mean and that of thermal distribution equals the mean square.
Thus, if the threshold is well-above the mean photon number of the signal, the detection loses most of the signal, and if it is well below the mean photon number, it is contaminated with noise without gaining signal.  

As seen in Fig. \ref{fig2}, the quantum SNR does not always exceed the classical SNR. Figure \ref{fig:3b} is a parameter-space plot, showing the parameters under which quantum detection is superior. Below the line (the darker area) threshold detection presents better SNR. 
As expected from Eq. \ref{Eq:therMonoton}, the area, where quantum detection outperforms the classical detection, grows as the threshold number is increased. We note that the curved point of each graph holds $N\approx \bar{n}_{\rm th}$. This fact may help to set the threshold as in most applications the noise intensity is approximately known or can be easily measured. 

In the same manner, it seems from the right bottom side of Fig. \ref{fig:3b} that threshold detection always gives better results where the noise is high and the signal is low. Thus, in high-noise low-signal regime, threshold detection is definitely preferable. 

We note that the average photon numbers ($\bar{n}_{\rm p},\,\bar{n}_{\rm th}$) are the measured averages, i.e. it already accounts for the loss of the detector. Other effects of the PNRD were considered, based on our PNRD model \cite{cohen2018absolute}, and  those effects  changed  the  results  slightly. In particular, nonlinear loss has low effect on the results, because we limited our signal to a few photons where the nonlinear loss is negligible (see Fig. S2 and the discussion in the supplementary \cite{WinNT}). 

\begin{figure}[tb]
\centering
   \begin{subfigure}[b]{1\columnwidth}
   \includegraphics[angle=0,width=1\columnwidth]{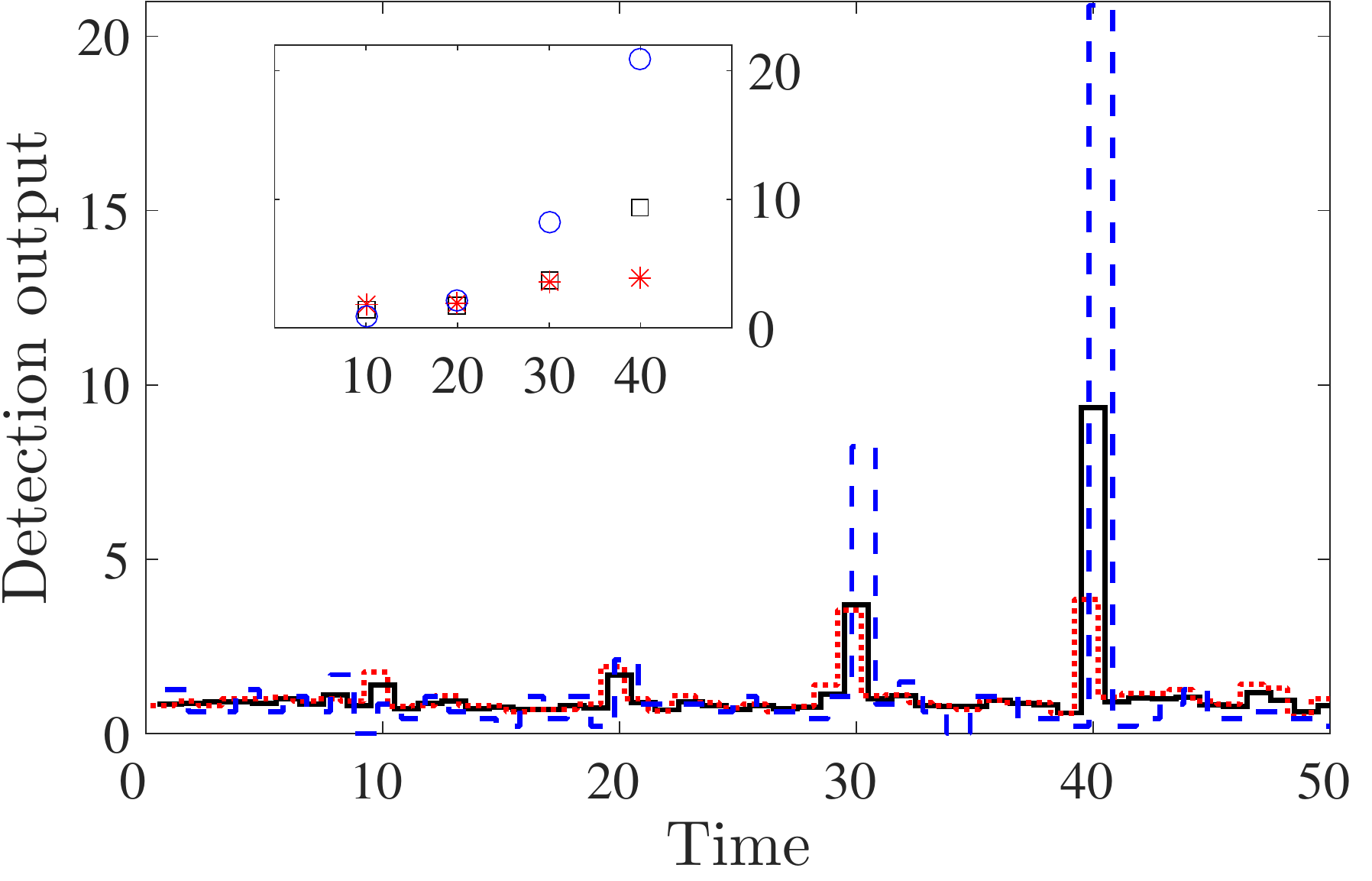}
   \caption{}    
   \label{fig:4a}
\end{subfigure}
\begin{subfigure}[b]{1\columnwidth}
   \includegraphics[angle=0,width=1\columnwidth]{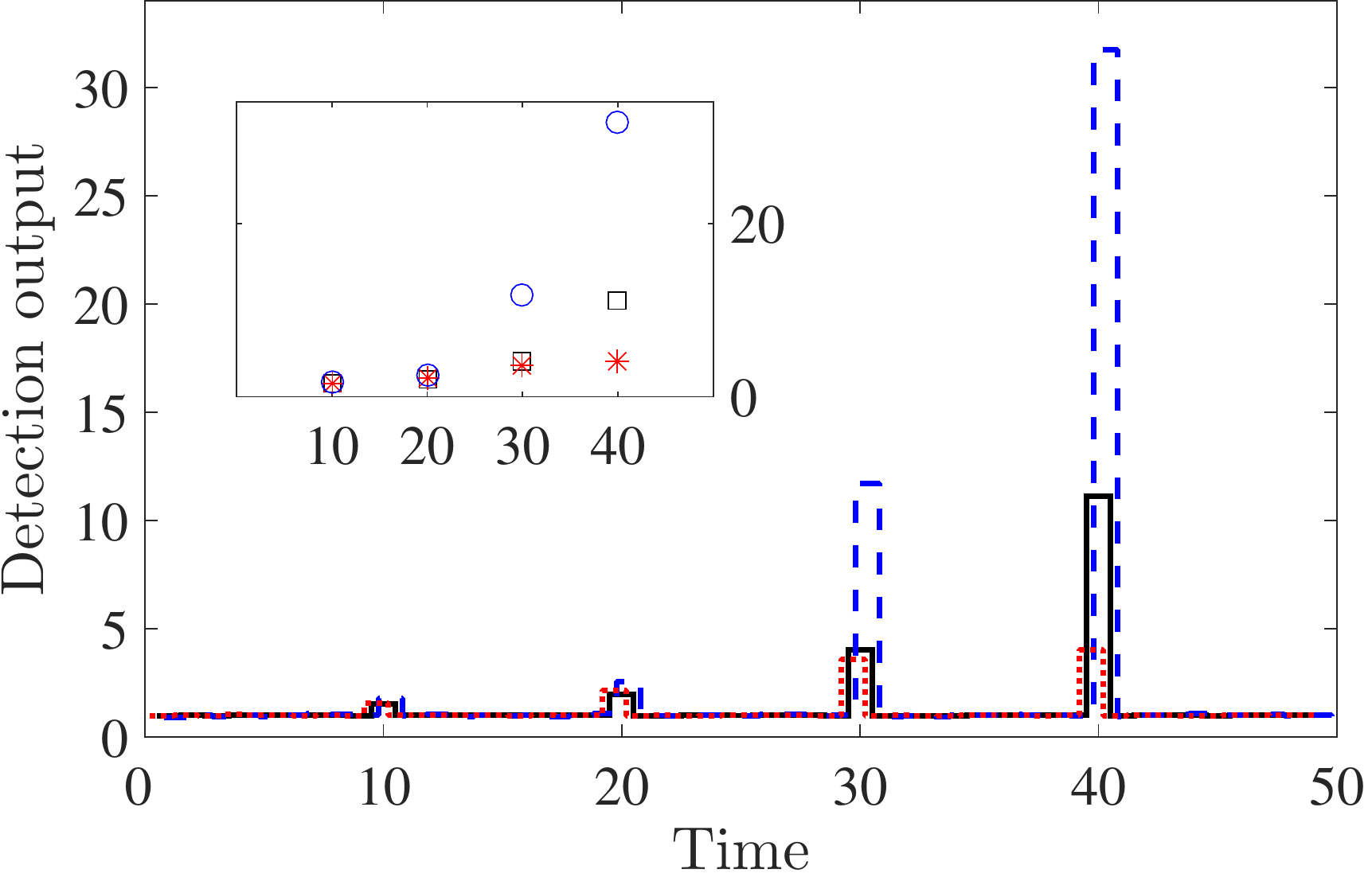}
   \caption{}
   \label{fig:4b}
\end{subfigure}
\caption{The simulation results comparing intensity detection and thresholding detection for 100 \textbf{(a)} and 10,000 \textbf{(b)} repetitions. The intensity detection is plotted with solid black line, two-photon thresholding with red dotted line, and five-photon thresholding with blue dashed line. The three graphs are slightly shifted, for visual purposes. The signal height is normalized such that the noise average is one. The inset shows the same comparison only for the time bins with the coherent photons. The intensity detection is plotted with black boxes, two-photon thresholding with red asterisks, and five-photon thresholding with blue circles.} \label{fig4}
\end{figure}

While Eq. \ref{Eq:quSNR} and Fig. \ref{fig2} show the average results for the quantum SNR and SNR ratio (i.e. infinite ensemble of measurement samplings), most applications may sample the signal only a few times. We simulate multi-target range-finding to show the improvement with a finite number of samplings. In the simulation, the time is divided to 50 time-bins, where the thermal noise is fixed with $\bar n_{\rm th} = 1$. Each time-bin contains noise photons distributed thermally. Four targets are simulated by adding photons with a Poisson distribution of $0.5\,,1\,,3$ and $10$ mean photon numbers at times of $10,20,30$ and $40$, respectively. The simulation runs 100 and 10,000 times, where the former is equivalent to less-than-a-second operation of a typical rangefinder. 

The simulation results are shown in Fig. \ref{fig4}. Naturally, the effect of low sampling is larger fluctuations, which can be seen in Fig. \ref{fig:4a}, especially for five-photon thresholding where the detection rate is low. The weak target with $\bar{n}_{\rm p}=0.5$ is detected well with two-photon thresholding but not detected at all with five-photon thresholding. This effect is again due to the detection rate. When the number of simulation repetitions is increased, the ratio of the SNR approaches the values of Fig. \ref{fig2}. For the target with $\bar{n}_{\rm p}=10$, the output of five-photon thresholding is 31.7 and of intensity is 11.1. As the noise is normalized to one, the ratio of the SNR is just $\frac{31.7}{11.1}=2.86$ which is exactly the result of Fig. \ref{fig2}. For the weak target with $\bar{n}_{\rm p}=0.5$, the output of two-photon thresholding is 1.58, of five-photon thresholding is 1.77 and of intensity is 1.51, which gives SNR ratio of 1.04 and 1.17 where 1.05 and 1.10 are deduced from Fig. \ref{fig2}.

We propose to implement the threshold detector with PNRD. There may be other implementation methods, such as N-photon-ionization processes.  Additionally, other detection protocols using PNRDs may give higher gain of the localized photon distribution, and thus, better SNR improvement. Examples include exact photon-number detection (i.e. projecting on a specific Fock state) \cite{khoury2006nonlinear} and a range of photon-number detection. These protocols require knowledge about the signal intensity and are suited to applications with known signal intensity. Threshold detection does not require knowledge about the signal intensity, and thus is suited to applications like range-finding and LIDAR, where the signal intensity is a priory unknown.

\textit{Summary.}\----We have shown that PNRDs can provide better SNR by thresholding the photon number, instead of directly detecting intensity. Additionally, we have theoretically tested our results for imperfect PNRD, including but not only nonlinear loss. This leads to a slightly lower SNR. The method seems to always improve the SNR in the high-noise low-signal regime.
The method has been implemented in rangefinders and LIDARs, but can also be used for any application with low-signal detection in the presence of thermal noise.

\textit{Acknowledgements.}\----LC, ESM, and JPD would like to acknowledge support from the Air Force Office of Scientific Research, the Army Research Office, the Defense Advanced Research Projects Agency, and the National Science Foundation.

\nocite{dovrat2012simulations}
\nocite{fox2006quantum}
\bibliography{sample}

\begin{thebibliography}{23}%
\makeatletter
\providecommand \@ifxundefined [1]{%
 \@ifx{#1\undefined}
}%
\providecommand \@ifnum [1]{%
 \ifnum #1\expandafter \@firstoftwo
 \else \expandafter \@secondoftwo
 \fi
}%
\providecommand \@ifx [1]{%
 \ifx #1\expandafter \@firstoftwo
 \else \expandafter \@secondoftwo
 \fi
}%
\providecommand \natexlab [1]{#1}%
\providecommand \enquote  [1]{``#1''}%
\providecommand \bibnamefont  [1]{#1}%
\providecommand \bibfnamefont [1]{#1}%
\providecommand \citenamefont [1]{#1}%
\providecommand \href@noop [0]{\@secondoftwo}%
\providecommand \href [0]{\begingroup \@sanitize@url \@href}%
\providecommand \@href[1]{\@@startlink{#1}\@@href}%
\providecommand \@@href[1]{\endgroup#1\@@endlink}%
\providecommand \@sanitize@url [0]{\catcode `\\12\catcode `\$12\catcode
  `\&12\catcode `\#12\catcode `\^12\catcode `\_12\catcode `\%12\relax}%
\providecommand \@@startlink[1]{}%
\providecommand \@@endlink[0]{}%
\providecommand \url  [0]{\begingroup\@sanitize@url \@url }%
\providecommand \@url [1]{\endgroup\@href {#1}{\urlprefix }}%
\providecommand \urlprefix  [0]{URL }%
\providecommand \Eprint [0]{\href }%
\providecommand \doibase [0]{http://dx.doi.org/}%
\providecommand \selectlanguage [0]{\@gobble}%
\providecommand \bibinfo  [0]{\@secondoftwo}%
\providecommand \bibfield  [0]{\@secondoftwo}%
\providecommand \translation [1]{[#1]}%
\providecommand \BibitemOpen [0]{}%
\providecommand \bibitemStop [0]{}%
\providecommand \bibitemNoStop [0]{.\EOS\space}%
\providecommand \EOS [0]{\spacefactor3000\relax}%
\providecommand \BibitemShut  [1]{\csname bibitem#1\endcsname}%
\let\auto@bib@innerbib\@empty
\bibitem [{\citenamefont {Warburton}\ \emph {et~al.}(2007)\citenamefont
  {Warburton}, \citenamefont {McCarthy}, \citenamefont {Wallace}, \citenamefont
  {Hernandez-Marin}, \citenamefont {Hadfield}, \citenamefont {Nam},\ and\
  \citenamefont {Buller}}]{warburton2007subcentimeter}%
  \BibitemOpen
  \bibfield  {author} {\bibinfo {author} {\bibfnamefont {R.~E.}\ \bibnamefont
  {Warburton}}, \bibinfo {author} {\bibfnamefont {A.}~\bibnamefont {McCarthy}},
  \bibinfo {author} {\bibfnamefont {A.~M.}\ \bibnamefont {Wallace}}, \bibinfo
  {author} {\bibfnamefont {S.}~\bibnamefont {Hernandez-Marin}}, \bibinfo
  {author} {\bibfnamefont {R.~H.}\ \bibnamefont {Hadfield}}, \bibinfo {author}
  {\bibfnamefont {S.~W.}\ \bibnamefont {Nam}}, \ and\ \bibinfo {author}
  {\bibfnamefont {G.~S.}\ \bibnamefont {Buller}},\ }\href@noop {} {\bibfield
  {journal} {\bibinfo  {journal} {Optics letters}\ }\textbf {\bibinfo {volume}
  {32}},\ \bibinfo {pages} {2266} (\bibinfo {year} {2007})}\BibitemShut
  {NoStop}%
\bibitem [{\citenamefont {Howland}\ \emph {et~al.}(2013)\citenamefont
  {Howland}, \citenamefont {Lum}, \citenamefont {Ware},\ and\ \citenamefont
  {Howell}}]{howland2013photon}%
  \BibitemOpen
  \bibfield  {author} {\bibinfo {author} {\bibfnamefont {G.~A.}\ \bibnamefont
  {Howland}}, \bibinfo {author} {\bibfnamefont {D.~J.}\ \bibnamefont {Lum}},
  \bibinfo {author} {\bibfnamefont {M.~R.}\ \bibnamefont {Ware}}, \ and\
  \bibinfo {author} {\bibfnamefont {J.~C.}\ \bibnamefont {Howell}},\
  }\href@noop {} {\bibfield  {journal} {\bibinfo  {journal} {Optics express}\
  }\textbf {\bibinfo {volume} {21}},\ \bibinfo {pages} {23822} (\bibinfo {year}
  {2013})}\BibitemShut {NoStop}%
\bibitem [{\citenamefont {Pawlikowska}\ \emph {et~al.}(2017)\citenamefont
  {Pawlikowska}, \citenamefont {Halimi}, \citenamefont {Lamb},\ and\
  \citenamefont {Buller}}]{pawlikowska2017single}%
  \BibitemOpen
  \bibfield  {author} {\bibinfo {author} {\bibfnamefont {A.~M.}\ \bibnamefont
  {Pawlikowska}}, \bibinfo {author} {\bibfnamefont {A.}~\bibnamefont {Halimi}},
  \bibinfo {author} {\bibfnamefont {R.~A.}\ \bibnamefont {Lamb}}, \ and\
  \bibinfo {author} {\bibfnamefont {G.~S.}\ \bibnamefont {Buller}},\
  }\href@noop {} {\bibfield  {journal} {\bibinfo  {journal} {Optics express}\
  }\textbf {\bibinfo {volume} {25}},\ \bibinfo {pages} {11919} (\bibinfo {year}
  {2017})}\BibitemShut {NoStop}%
\bibitem [{\citenamefont {Li}\ \emph {et~al.}(2019)\citenamefont {Li},
  \citenamefont {Huang}, \citenamefont {Cao}, \citenamefont {Wang},
  \citenamefont {Li}, \citenamefont {Jin}, \citenamefont {Yu}, \citenamefont
  {Zhang}, \citenamefont {Zhang}, \citenamefont {Peng} \emph
  {et~al.}}]{li2019single}%
  \BibitemOpen
  \bibfield  {author} {\bibinfo {author} {\bibfnamefont {Z.-P.}\ \bibnamefont
  {Li}}, \bibinfo {author} {\bibfnamefont {X.}~\bibnamefont {Huang}}, \bibinfo
  {author} {\bibfnamefont {Y.}~\bibnamefont {Cao}}, \bibinfo {author}
  {\bibfnamefont {B.}~\bibnamefont {Wang}}, \bibinfo {author} {\bibfnamefont
  {Y.-H.}\ \bibnamefont {Li}}, \bibinfo {author} {\bibfnamefont
  {W.}~\bibnamefont {Jin}}, \bibinfo {author} {\bibfnamefont {C.}~\bibnamefont
  {Yu}}, \bibinfo {author} {\bibfnamefont {J.}~\bibnamefont {Zhang}}, \bibinfo
  {author} {\bibfnamefont {Q.}~\bibnamefont {Zhang}}, \bibinfo {author}
  {\bibfnamefont {C.-Z.}\ \bibnamefont {Peng}},  \emph {et~al.},\ }\href@noop
  {} {\bibfield  {journal} {\bibinfo  {journal} {arXiv preprint
  arXiv:1904.10341}\ } (\bibinfo {year} {2019})}\BibitemShut {NoStop}%
\bibitem [{\citenamefont {Bao}\ \emph {et~al.}(2014)\citenamefont {Bao},
  \citenamefont {Liang}, \citenamefont {Wang}, \citenamefont {Li},
  \citenamefont {Wu}, \citenamefont {Wu},\ and\ \citenamefont
  {Zeng}}]{bao2014laser}%
  \BibitemOpen
  \bibfield  {author} {\bibinfo {author} {\bibfnamefont {Z.}~\bibnamefont
  {Bao}}, \bibinfo {author} {\bibfnamefont {Y.}~\bibnamefont {Liang}}, \bibinfo
  {author} {\bibfnamefont {Z.}~\bibnamefont {Wang}}, \bibinfo {author}
  {\bibfnamefont {Z.}~\bibnamefont {Li}}, \bibinfo {author} {\bibfnamefont
  {E.}~\bibnamefont {Wu}}, \bibinfo {author} {\bibfnamefont {G.}~\bibnamefont
  {Wu}}, \ and\ \bibinfo {author} {\bibfnamefont {H.}~\bibnamefont {Zeng}},\
  }\href@noop {} {\bibfield  {journal} {\bibinfo  {journal} {Applied optics}\
  }\textbf {\bibinfo {volume} {53}},\ \bibinfo {pages} {3908} (\bibinfo {year}
  {2014})}\BibitemShut {NoStop}%
\bibitem [{\citenamefont {Sher}\ \emph {et~al.}(2018)\citenamefont {Sher},
  \citenamefont {Cohen}, \citenamefont {Istrati},\ and\ \citenamefont
  {Eisenberg}}]{sher2018low}%
  \BibitemOpen
  \bibfield  {author} {\bibinfo {author} {\bibfnamefont {Y.}~\bibnamefont
  {Sher}}, \bibinfo {author} {\bibfnamefont {L.}~\bibnamefont {Cohen}},
  \bibinfo {author} {\bibfnamefont {D.}~\bibnamefont {Istrati}}, \ and\
  \bibinfo {author} {\bibfnamefont {H.~S.}\ \bibnamefont {Eisenberg}},\ }in\
  \href@noop {} {\emph {\bibinfo {booktitle} {Emerging Digital Micromirror
  Device Based Systems and Applications X}}},\ Vol.\ \bibinfo {volume} {10546}\
  (\bibinfo {organization} {International Society for Optics and Photonics},\
  \bibinfo {year} {2018})\ p.\ \bibinfo {pages} {105460J}\BibitemShut {NoStop}%
\bibitem [{\citenamefont {Dorner}\ \emph {et~al.}(2009)\citenamefont {Dorner},
  \citenamefont {Demkowicz-Dobrzanski}, \citenamefont {Smith}, \citenamefont
  {Lundeen}, \citenamefont {Wasilewski}, \citenamefont {Banaszek},\ and\
  \citenamefont {Walmsley}}]{dorner2009optimal}%
  \BibitemOpen
  \bibfield  {author} {\bibinfo {author} {\bibfnamefont {U.}~\bibnamefont
  {Dorner}}, \bibinfo {author} {\bibfnamefont {R.}~\bibnamefont
  {Demkowicz-Dobrzanski}}, \bibinfo {author} {\bibfnamefont {B.}~\bibnamefont
  {Smith}}, \bibinfo {author} {\bibfnamefont {J.}~\bibnamefont {Lundeen}},
  \bibinfo {author} {\bibfnamefont {W.}~\bibnamefont {Wasilewski}}, \bibinfo
  {author} {\bibfnamefont {K.}~\bibnamefont {Banaszek}}, \ and\ \bibinfo
  {author} {\bibfnamefont {I.}~\bibnamefont {Walmsley}},\ }\href@noop {}
  {\bibfield  {journal} {\bibinfo  {journal} {Physical review letters}\
  }\textbf {\bibinfo {volume} {102}},\ \bibinfo {pages} {040403} (\bibinfo
  {year} {2009})}\BibitemShut {NoStop}%
\bibitem [{\citenamefont {Lee}\ \emph {et~al.}(2009)\citenamefont {Lee},
  \citenamefont {Huver}, \citenamefont {Lee}, \citenamefont {Kaplan},
  \citenamefont {McCracken}, \citenamefont {Min}, \citenamefont {Uskov},
  \citenamefont {Wildfeuer}, \citenamefont {Veronis},\ and\ \citenamefont
  {Dowling}}]{lee2009optimization}%
  \BibitemOpen
  \bibfield  {author} {\bibinfo {author} {\bibfnamefont {T.-W.}\ \bibnamefont
  {Lee}}, \bibinfo {author} {\bibfnamefont {S.~D.}\ \bibnamefont {Huver}},
  \bibinfo {author} {\bibfnamefont {H.}~\bibnamefont {Lee}}, \bibinfo {author}
  {\bibfnamefont {L.}~\bibnamefont {Kaplan}}, \bibinfo {author} {\bibfnamefont
  {S.~B.}\ \bibnamefont {McCracken}}, \bibinfo {author} {\bibfnamefont
  {C.}~\bibnamefont {Min}}, \bibinfo {author} {\bibfnamefont {D.~B.}\
  \bibnamefont {Uskov}}, \bibinfo {author} {\bibfnamefont {C.~F.}\ \bibnamefont
  {Wildfeuer}}, \bibinfo {author} {\bibfnamefont {G.}~\bibnamefont {Veronis}},
  \ and\ \bibinfo {author} {\bibfnamefont {J.~P.}\ \bibnamefont {Dowling}},\
  }\href@noop {} {\bibfield  {journal} {\bibinfo  {journal} {Physical Review
  A}\ }\textbf {\bibinfo {volume} {80}},\ \bibinfo {pages} {063803} (\bibinfo
  {year} {2009})}\BibitemShut {NoStop}%
\bibitem [{\citenamefont {Jiang}\ \emph {et~al.}(2013)\citenamefont {Jiang},
  \citenamefont {Lee}, \citenamefont {Gerry},\ and\ \citenamefont
  {Dowling}}]{jiang2013super}%
  \BibitemOpen
  \bibfield  {author} {\bibinfo {author} {\bibfnamefont {K.}~\bibnamefont
  {Jiang}}, \bibinfo {author} {\bibfnamefont {H.}~\bibnamefont {Lee}}, \bibinfo
  {author} {\bibfnamefont {C.~C.}\ \bibnamefont {Gerry}}, \ and\ \bibinfo
  {author} {\bibfnamefont {J.~P.}\ \bibnamefont {Dowling}},\ }\href@noop {}
  {\bibfield  {journal} {\bibinfo  {journal} {Journal of Applied Physics}\
  }\textbf {\bibinfo {volume} {114}},\ \bibinfo {pages} {193102} (\bibinfo
  {year} {2013})}\BibitemShut {NoStop}%
\bibitem [{\citenamefont {Qian}\ \emph {et~al.}(2015)\citenamefont {Qian},
  \citenamefont {Little}, \citenamefont {Howell},\ and\ \citenamefont
  {Eberly}}]{qian2015shifting}%
  \BibitemOpen
  \bibfield  {author} {\bibinfo {author} {\bibfnamefont {X.-F.}\ \bibnamefont
  {Qian}}, \bibinfo {author} {\bibfnamefont {B.}~\bibnamefont {Little}},
  \bibinfo {author} {\bibfnamefont {J.~C.}\ \bibnamefont {Howell}}, \ and\
  \bibinfo {author} {\bibfnamefont {J.}~\bibnamefont {Eberly}},\ }\href@noop {}
  {\bibfield  {journal} {\bibinfo  {journal} {Optica}\ }\textbf {\bibinfo
  {volume} {2}},\ \bibinfo {pages} {611} (\bibinfo {year} {2015})}\BibitemShut
  {NoStop}%
\bibitem [{\citenamefont {Giustina}\ \emph {et~al.}(2013)\citenamefont
  {Giustina}, \citenamefont {Mech}, \citenamefont {Ramelow}, \citenamefont
  {Wittmann}, \citenamefont {Kofler}, \citenamefont {Beyer}, \citenamefont
  {Lita}, \citenamefont {Calkins}, \citenamefont {Gerrits}, \citenamefont {Nam}
  \emph {et~al.}}]{giustina2013bell}%
  \BibitemOpen
  \bibfield  {author} {\bibinfo {author} {\bibfnamefont {M.}~\bibnamefont
  {Giustina}}, \bibinfo {author} {\bibfnamefont {A.}~\bibnamefont {Mech}},
  \bibinfo {author} {\bibfnamefont {S.}~\bibnamefont {Ramelow}}, \bibinfo
  {author} {\bibfnamefont {B.}~\bibnamefont {Wittmann}}, \bibinfo {author}
  {\bibfnamefont {J.}~\bibnamefont {Kofler}}, \bibinfo {author} {\bibfnamefont
  {J.}~\bibnamefont {Beyer}}, \bibinfo {author} {\bibfnamefont
  {A.}~\bibnamefont {Lita}}, \bibinfo {author} {\bibfnamefont {B.}~\bibnamefont
  {Calkins}}, \bibinfo {author} {\bibfnamefont {T.}~\bibnamefont {Gerrits}},
  \bibinfo {author} {\bibfnamefont {S.~W.}\ \bibnamefont {Nam}},  \emph
  {et~al.},\ }\href@noop {} {\bibfield  {journal} {\bibinfo  {journal}
  {Nature}\ }\textbf {\bibinfo {volume} {497}},\ \bibinfo {pages} {227}
  (\bibinfo {year} {2013})}\BibitemShut {NoStop}%
\bibitem [{\citenamefont {Bell}(1987)}]{bell1987speakable}%
  \BibitemOpen
  \bibfield  {author} {\bibinfo {author} {\bibfnamefont {J.~S.}\ \bibnamefont
  {Bell}},\ }\href@noop {} {\bibfield  {journal} {\bibinfo  {journal}
  {Cambridge University}\ } (\bibinfo {year} {1987})}\BibitemShut {NoStop}%
\bibitem [{\citenamefont {Busck}\ and\ \citenamefont
  {Heiselberg}(2004)}]{busck2004gated}%
  \BibitemOpen
  \bibfield  {author} {\bibinfo {author} {\bibfnamefont {J.}~\bibnamefont
  {Busck}}\ and\ \bibinfo {author} {\bibfnamefont {H.}~\bibnamefont
  {Heiselberg}},\ }\href@noop {} {\bibfield  {journal} {\bibinfo  {journal}
  {Applied optics}\ }\textbf {\bibinfo {volume} {43}},\ \bibinfo {pages} {4705}
  (\bibinfo {year} {2004})}\BibitemShut {NoStop}%
\bibitem [{\citenamefont {Dovrat}\ \emph
  {et~al.}(2012{\natexlab{a}})\citenamefont {Dovrat}, \citenamefont {Bakstein},
  \citenamefont {Istrati}, \citenamefont {Shaham},\ and\ \citenamefont
  {Eisenberg}}]{dovrat2012measurements}%
  \BibitemOpen
  \bibfield  {author} {\bibinfo {author} {\bibfnamefont {L.}~\bibnamefont
  {Dovrat}}, \bibinfo {author} {\bibfnamefont {M.}~\bibnamefont {Bakstein}},
  \bibinfo {author} {\bibfnamefont {D.}~\bibnamefont {Istrati}}, \bibinfo
  {author} {\bibfnamefont {A.}~\bibnamefont {Shaham}}, \ and\ \bibinfo {author}
  {\bibfnamefont {H.~S.}\ \bibnamefont {Eisenberg}},\ }\href@noop {} {\bibfield
   {journal} {\bibinfo  {journal} {Optics express}\ }\textbf {\bibinfo {volume}
  {20}},\ \bibinfo {pages} {2266} (\bibinfo {year}
  {2012}{\natexlab{a}})}\BibitemShut {NoStop}%
\bibitem [{\citenamefont {Aasi}\ \emph {et~al.}(2013)\citenamefont {Aasi},
  \citenamefont {Abadie}, \citenamefont {Abbott}, \citenamefont {Abbott},
  \citenamefont {Abbott}, \citenamefont {Abernathy}, \citenamefont {Adams},
  \citenamefont {Adams}, \citenamefont {Addesso}, \citenamefont {Adhikari}
  \emph {et~al.}}]{aasi2013enhanced}%
  \BibitemOpen
  \bibfield  {author} {\bibinfo {author} {\bibfnamefont {J.}~\bibnamefont
  {Aasi}}, \bibinfo {author} {\bibfnamefont {J.}~\bibnamefont {Abadie}},
  \bibinfo {author} {\bibfnamefont {B.}~\bibnamefont {Abbott}}, \bibinfo
  {author} {\bibfnamefont {R.}~\bibnamefont {Abbott}}, \bibinfo {author}
  {\bibfnamefont {T.}~\bibnamefont {Abbott}}, \bibinfo {author} {\bibfnamefont
  {M.}~\bibnamefont {Abernathy}}, \bibinfo {author} {\bibfnamefont
  {C.}~\bibnamefont {Adams}}, \bibinfo {author} {\bibfnamefont
  {T.}~\bibnamefont {Adams}}, \bibinfo {author} {\bibfnamefont
  {P.}~\bibnamefont {Addesso}}, \bibinfo {author} {\bibfnamefont
  {R.}~\bibnamefont {Adhikari}},  \emph {et~al.},\ }\href@noop {} {\bibfield
  {journal} {\bibinfo  {journal} {Nature Photonics}\ }\textbf {\bibinfo
  {volume} {7}},\ \bibinfo {pages} {613} (\bibinfo {year} {2013})}\BibitemShut
  {NoStop}%
\bibitem [{\citenamefont {Israel}\ \emph {et~al.}(2014)\citenamefont {Israel},
  \citenamefont {Rosen},\ and\ \citenamefont
  {Silberberg}}]{israel2014supersensitive}%
  \BibitemOpen
  \bibfield  {author} {\bibinfo {author} {\bibfnamefont {Y.}~\bibnamefont
  {Israel}}, \bibinfo {author} {\bibfnamefont {S.}~\bibnamefont {Rosen}}, \
  and\ \bibinfo {author} {\bibfnamefont {Y.}~\bibnamefont {Silberberg}},\
  }\href@noop {} {\bibfield  {journal} {\bibinfo  {journal} {Physical review
  letters}\ }\textbf {\bibinfo {volume} {112}},\ \bibinfo {pages} {103604}
  (\bibinfo {year} {2014})}\BibitemShut {NoStop}%
\bibitem [{\citenamefont {Escher}\ \emph {et~al.}(2011)\citenamefont {Escher},
  \citenamefont {de~Matos~Filho},\ and\ \citenamefont
  {Davidovich}}]{escher2011general}%
  \BibitemOpen
  \bibfield  {author} {\bibinfo {author} {\bibfnamefont {B.}~\bibnamefont
  {Escher}}, \bibinfo {author} {\bibfnamefont {R.}~\bibnamefont
  {de~Matos~Filho}}, \ and\ \bibinfo {author} {\bibfnamefont {L.}~\bibnamefont
  {Davidovich}},\ }\href@noop {} {\bibfield  {journal} {\bibinfo  {journal}
  {Nature Physics}\ }\textbf {\bibinfo {volume} {7}},\ \bibinfo {pages} {406}
  (\bibinfo {year} {2011})}\BibitemShut {NoStop}%
\bibitem [{\citenamefont {Cohen}\ \emph {et~al.}(2016)\citenamefont {Cohen},
  \citenamefont {Pilnyak}, \citenamefont {Istrati}, \citenamefont {Retzker},\
  and\ \citenamefont {Eisenberg}}]{cohen2016demonstration}%
  \BibitemOpen
  \bibfield  {author} {\bibinfo {author} {\bibfnamefont {L.}~\bibnamefont
  {Cohen}}, \bibinfo {author} {\bibfnamefont {Y.}~\bibnamefont {Pilnyak}},
  \bibinfo {author} {\bibfnamefont {D.}~\bibnamefont {Istrati}}, \bibinfo
  {author} {\bibfnamefont {A.}~\bibnamefont {Retzker}}, \ and\ \bibinfo
  {author} {\bibfnamefont {H.~S.}\ \bibnamefont {Eisenberg}},\ }\href@noop {}
  {\bibfield  {journal} {\bibinfo  {journal} {Physical Review A}\ }\textbf
  {\bibinfo {volume} {94}},\ \bibinfo {pages} {012324} (\bibinfo {year}
  {2016})}\BibitemShut {NoStop}%
\bibitem [{Win()}]{WinNT}%
  \BibitemOpen
  \href@noop {} {}\bibinfo {note} {See Supplemental Material [url] which
  includes Refs. [22-23].}\BibitemShut {Stop}%
\bibitem [{\citenamefont {Cohen}\ \emph {et~al.}(2018)\citenamefont {Cohen},
  \citenamefont {Pilnyak}, \citenamefont {Istrati}, \citenamefont {Studer},
  \citenamefont {Dowling},\ and\ \citenamefont
  {Eisenberg}}]{cohen2018absolute}%
  \BibitemOpen
  \bibfield  {author} {\bibinfo {author} {\bibfnamefont {L.}~\bibnamefont
  {Cohen}}, \bibinfo {author} {\bibfnamefont {Y.}~\bibnamefont {Pilnyak}},
  \bibinfo {author} {\bibfnamefont {D.}~\bibnamefont {Istrati}}, \bibinfo
  {author} {\bibfnamefont {N.~M.}\ \bibnamefont {Studer}}, \bibinfo {author}
  {\bibfnamefont {J.~P.}\ \bibnamefont {Dowling}}, \ and\ \bibinfo {author}
  {\bibfnamefont {H.~S.}\ \bibnamefont {Eisenberg}},\ }\href@noop {} {\bibfield
   {journal} {\bibinfo  {journal} {Physical Review A}\ }\textbf {\bibinfo
  {volume} {98}},\ \bibinfo {pages} {013811} (\bibinfo {year}
  {2018})}\BibitemShut {NoStop}%
\bibitem [{\citenamefont {Khoury}\ \emph {et~al.}(2006)\citenamefont {Khoury},
  \citenamefont {Eisenberg}, \citenamefont {Fonseca},\ and\ \citenamefont
  {Bouwmeester}}]{khoury2006nonlinear}%
  \BibitemOpen
  \bibfield  {author} {\bibinfo {author} {\bibfnamefont {G.}~\bibnamefont
  {Khoury}}, \bibinfo {author} {\bibfnamefont {H.~S.}\ \bibnamefont
  {Eisenberg}}, \bibinfo {author} {\bibfnamefont {E.}~\bibnamefont {Fonseca}},
  \ and\ \bibinfo {author} {\bibfnamefont {D.}~\bibnamefont {Bouwmeester}},\
  }\href@noop {} {\bibfield  {journal} {\bibinfo  {journal} {Physical review
  letters}\ }\textbf {\bibinfo {volume} {96}},\ \bibinfo {pages} {203601}
  (\bibinfo {year} {2006})}\BibitemShut {NoStop}%
\bibitem [{\citenamefont {Dovrat}\ \emph
  {et~al.}(2012{\natexlab{b}})\citenamefont {Dovrat}, \citenamefont {Bakstein},
  \citenamefont {Istrati},\ and\ \citenamefont
  {Eisenberg}}]{dovrat2012simulations}%
  \BibitemOpen
  \bibfield  {author} {\bibinfo {author} {\bibfnamefont {L.}~\bibnamefont
  {Dovrat}}, \bibinfo {author} {\bibfnamefont {M.}~\bibnamefont {Bakstein}},
  \bibinfo {author} {\bibfnamefont {D.}~\bibnamefont {Istrati}}, \ and\
  \bibinfo {author} {\bibfnamefont {H.~S.}\ \bibnamefont {Eisenberg}},\
  }\href@noop {} {\bibfield  {journal} {\bibinfo  {journal} {Physica Scripta}\
  }\textbf {\bibinfo {volume} {2012}},\ \bibinfo {pages} {014010} (\bibinfo
  {year} {2012}{\natexlab{b}})}\BibitemShut {NoStop}%
\bibitem [{\citenamefont {Fox}(2006)}]{fox2006quantum}%
  \BibitemOpen
  \bibfield  {author} {\bibinfo {author} {\bibfnamefont {M.}~\bibnamefont
  {Fox}},\ }\href@noop {} {\emph {\bibinfo {title} {Quantum optics: an
  introduction}}},\ Vol.~\bibinfo {volume} {15}\ (\bibinfo  {publisher} {OUP
  Oxford},\ \bibinfo {year} {2006})\BibitemShut {NoStop}%
\end{thebibliography}%


%


\end{document}